\documentclass[prl, reprint,showpacs,superscriptaddress,longbibliography]{revtex4-2}
 
\usepackage{graphicx}
\usepackage{dcolumn}
\usepackage{graphicx}
\usepackage{epsfig}
\usepackage{epsf}
\usepackage{amssymb,amsmath,amsthm,bm,dsfont}
\usepackage{empheq,cases}
\usepackage{mathtools}
\usepackage{multirow}
\usepackage[colorlinks=true,linkcolor=blue!50!black,citecolor=blue,urlcolor=blue,pdfauthor={ },pdftitle={ },pdfsubject={ },pdfkeywords={ }]{hyperref}
\usepackage{tikz}
\usepackage{pgfplots}
\usepackage{xcolor}
\usepackage{lipsum}

\usepgfplotslibrary{patchplots}
\usetikzlibrary{decorations.text }
\usetikzlibrary{shapes.symbols}
\usetikzlibrary{intersections,shapes.arrows}
\usetikzlibrary{decorations.markings}
\usetikzlibrary{mindmap,trees, backgrounds,fadings,shadows}
\usetikzlibrary{positioning,calc}
\usetikzlibrary{arrows,snakes,shapes}
\usetikzlibrary{patterns}

\definecolor{c1}{RGB}{219,68,56}
\definecolor{c2}{RGB}{74,91,163}
\definecolor{c3}{RGB}{105,182,106}

\theoremstyle{definition}

\definecolor{c1}{RGB}{219,68,56}
\definecolor{c2}{RGB}{74,91,163}
\definecolor{c3}{RGB}{105,182,106}

\usepackage{multirow}
\usepackage{dcolumn}
\newcolumntype{d}[1]{D{.}{.}{#1}}

\usepackage[colorinlistoftodos,bordercolor=blue!50,backgroundcolor=blue!10,linecolor=blue!50,textsize=tiny]{todonotes}
\usepackage{changes}
\setaddedmarkup{\color{red} \uline{#1}}
\setdeletedmarkup{\color{red} \sout{#1}}
\sethighlightmarkup{{\color{blue} #1}}

\usepackage{times}

\begin{document}

\title{Quantum Control for Time-dependent Noise by Inverse Geometric Optimization}

\author{Xiaodong Yang}
\affiliation{Shenzhen Institute for Quantum Science and Engineering, Southern University of Science and Technology, Shenzhen, 518055, China}
\affiliation{International Quantum Academy, Shenzhen, 518055, China}
\affiliation{Guangdong Provincial Key Laboratory of Quantum Science and Engineering, Southern University of Science and Technology, Shenzhen, 518055, China}

\author{Xinfang Nie}
\affiliation{Department of Physics, Southern University of Science and Technology, Shenzhen, 518055, China}
\affiliation{Shenzhen Institute for Quantum Science and Engineering, Southern University of Science and Technology, Shenzhen, 518055, China}
\affiliation{International Quantum Academy, Shenzhen, 518055, China}
\affiliation{Guangdong Provincial Key Laboratory of Quantum Science and Engineering, Southern University of Science and Technology, Shenzhen, 518055, China}

\author{Tao Xin}
\affiliation{Shenzhen Institute for Quantum Science and Engineering, Southern University of Science and Technology, Shenzhen, 518055, China}
\affiliation{International Quantum Academy, Shenzhen, 518055, China}

\affiliation{Guangdong Provincial Key Laboratory of Quantum Science and Engineering, Southern University of Science and Technology, Shenzhen, 518055, China}

\author{Dawei Lu}
\email{ludw@sustech.edu.cn}
\affiliation{Department of Physics, Southern University of Science and Technology, Shenzhen, 518055, China}
\affiliation{Shenzhen Institute for Quantum Science and Engineering, Southern University of Science and Technology, Shenzhen, 518055, China}
\affiliation{International Quantum Academy, Shenzhen, 518055, China}
\affiliation{Guangdong Provincial Key Laboratory of Quantum Science and Engineering, Southern University of Science and Technology, Shenzhen, 518055, China}

\author{Jun Li}
\email{lij3@sustech.edu.cn}
\affiliation{Shenzhen Institute for Quantum Science and Engineering, Southern University of Science and Technology, Shenzhen, 518055, China}
\affiliation{International Quantum Academy, Shenzhen, 518055, China}
\affiliation{Guangdong Provincial Key Laboratory of Quantum Science and Engineering, Southern University of Science and Technology, Shenzhen, 518055, China}

\begin{abstract}
Quantum systems are exceedingly difficult to engineer  because they are sensitive to various types of noises. In particular,   time-dependent noises are frequently encountered in   experiments but how to overcome them remains a challenging problem.
In this work, we extend  and apply  the recently proposed robust control technique of inverse geometric optimization to  time-dependent noises   by working it in  the filter-function  formalism.  The  basic idea  is to   parameterize the control filter function geometrically and   minimize its   overlap with the noise spectral density. This then effectively reduces  the noise susceptibility of the controlled system evolution.  
We show that the proposed method can produce high-quality  robust pulses for realizing  desired  quantum evolutions under   realistic  noise models, and thus will find practical applications for current physical platforms.
\end{abstract}

\maketitle

\begin{figure}[t] 
\includegraphics[width=\linewidth]{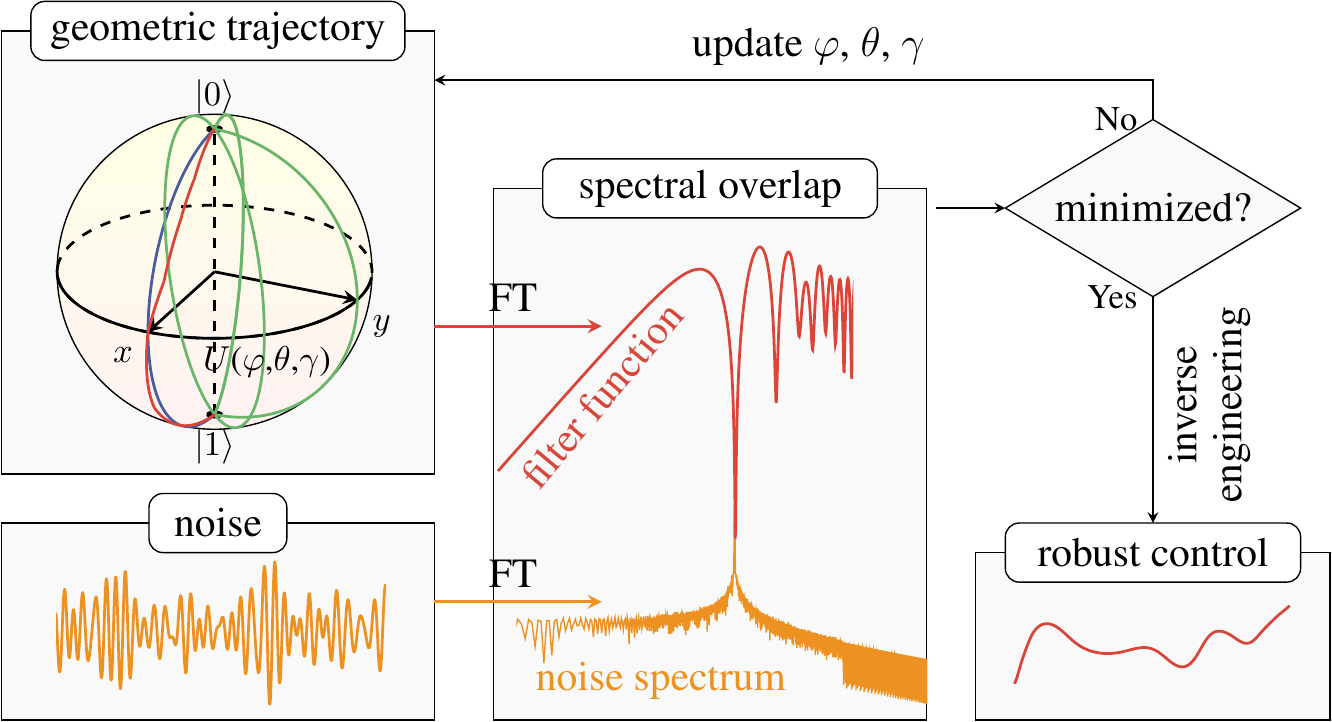}
\caption{Schematic diagram of the method of geometric- and FF- based    pulse optimization for resisting    time-dependent, stochastic noise. Since there exist many evolution trajectories realizing the same    target but with   different extent of noise filtering capabilities, the goal is hence to find     a noise-resilient     trajectory.  For   example, in the single-qubit case, the geometric trajectories generated by rectangular wave (blue line), composite pulse (green line), and a robust shaped pulse (red line) are plotted on the Bloch sphere for comparison, all implementing the same state transfer  $|0\rangle \to |1\rangle$.  Robust trajectory is found by minimizing the overlap of  its associated control filter function  with the noise spectral density. The pulse shape that generates this trajectory is then   obtained through inverse engineering. } 
\label{BlochPath}
\end{figure}

{\emph{Introduction.---}The ability to precisely manipulate quantum systems against    noise    is central to practical quantum information processing \cite{RevModPhys.88.041001}. There have been developed a variety of robust quantum control methods, such as  composite pulses \cite{Levitt86,PhysRevA.67.042308,PhysRevA.70.052318}, dynamical decoupling \cite{PhysRevLett.82.2417,VLK03,SAS12}, sampling-based learning control \cite{PhysRevA.89.023402,dong2015sampling},  geometric-formalism-based pulse control  \cite{PhysRevLett.111.050404,PhysRevLett.125.250403,BWS15,PhysRevA.99.052321,PRXQuantum.2.010341,PhysRevA.100.062310}. Many of these methods   assume   the considered noise to be   quasi-static, i.e., slow enough compared to the   operation time, which is however not always a valid noise model in reality. Actually, time-dependent noises are routinely encountered   in  experiments. For example, $1/f$ type noise, which contains wide distribution of correlation times   \cite{RevModPhys.86.361}, is present in  many solid-state qubit platforms  such as superconducting qubits \cite{clarke2008superconducting,Oliver19} and semiconductor quantum dots \cite{ladd2010quantum,PhysRevLett.118.177702}. Therefore, in order to further enhance experimental control fidelities, it is of vital importance to develop robust quantum control for general time-dependent noises.

%
%


Attempts to address errors induced by time-dependent  noises in   quantum system engineering are challenging. Results to date suggest  that  conventional  methods usually have their limitations. For example, composite pulses, originally designed to tackle static, systematic errors, were found to be  robust  to fluctuating noises up to as fast as around  10\% of the   Rabi frequency \cite{PhysRevA.90.012316}.
Dynamical decoupling can protect quantum coherence in  a fluctuating environment,   but it requires rapid and strong control modulation, which might be problematic to realize experimentally. Moreover, how to incorporate dynamical decoupling into the task of realizing arbitrary quantum operations is still  not fully clear \cite{PhysRevLett.112.050502}.  
Optimal control provides  a flexible and generically applicable approach, in which the requirements of pulse smoothness and  robustness   can   be added   as   optimization constraints \cite{WG07}. Usually, the control variables to be optimized are   temporal pulse parameters such as amplitudes and phases. Alternatively, optimization can  be done in the dynamical variable space with a geometric flavor, as proposed and developed in Refs. \cite{PhysRevLett.111.050404,PhysRevLett.125.250403,BWS15,PhysRevA.99.052321,PRXQuantum.2.010341,PhysRevA.100.062310}, yet only   static errors have been considered therein. 

In this work,   we consider  combing the geometric-based optimal control method with the filter function (FF) formalism \cite{PhysRevB.67.094510,PhysRevLett.93.130406,PhysRevLett.98.100504} to overcome these limitations for the purpose of resisting time-dependent noises.   FFs were originally introduced to evaluate    operational infidelities under   stationary stochastic noises, and have proven   very useful   in quantum control, especially for   designing  dynamical decoupling sequences \cite{PhysRevLett.101.010403,Biercuk09,PhysRevLett.103.040501,PhysRevLett.104.040401,Kabytayev15,biercuk2011dynamical,PhysRevLett.103.040501}. Recently, there have been studies on incorporating FF into gradient-based optimal control \cite{PhysRevApplied.17.024006}.  Here, we take the geometric approach, that is, we first parameterize the controlled system evolution trajectory with dynamical variables, which corresponds to a parameterized filter function in the frequency domain, and then minimize the overlap of the filter function   and the noise spectral density; see  Fig. \ref{BlochPath} for  an illustration of the basic idea.

{   We give test  examples of finding robust optimal control (ROC) pulses for producing target quantum gate and state transfer under realistic,  experimentally relevant  noise environments.} It is  found that our robust pulses outperform   typical composite pulses in that their resultant FFs are suppressed at the characteristic frequencies of the considered noises, thus having much improved control fidelities.  A separate section is devoted to treat the   case of Markovian   noise   based directly on   the Bloch equation, and the optimization results show that the   $T_1$ and $T_2$ limit can be surpassed in the quantum state transfer task.
Finally, discussions and implications are presented.}

\emph{Inverse geometric engineering.---}We consider a prototypical robust quantum control model, i.e., a resonantly controlled two-level system under time-dependent detuning noise and control amplitude noise. By convention, we parameterize the control field  as  $\Omega(t)[\cos\phi(t), \sin\phi(t)]$ ($t \in [0,T]$), with $\Omega(t)$ ($ |\Omega(t)| \le \Omega_{\max}$) being the pulsed Rabi frequency  and  $ \phi(t) \in [-\pi, \pi]$ the phase. Taking into account of noises, we  have the following resonant frame Hamiltonian
\begin{equation}\label{eq1}
	H(t) 
	= \Omega (1+\epsilon_a(t)) \left[ \cos \phi \frac{\sigma_x}{2}  + \sin \phi \frac{\sigma_y}{2} \right]  + \epsilon_d(t) \frac{\sigma_z}{2}, 
\end{equation}
where $\epsilon_a(t),\epsilon_d(t)$ represent fluctuating noises on control amplitude and detuning, respectively,  and we introduce $E_a(t) \equiv \Omega[ \cos \phi \sigma_x/2  + \sin \phi  \sigma_y/2 ]$ and $E_d\equiv  \sigma_z/2$ as their corresponding noise operators.
Physically, control amplitude noise is usually due to imperfect fabricated components, noisy electronics or varied fields \cite{PhysRevA.91.052306}, while detuning may originate from, e.g., random shifts in control driving frequency, or Overhauser effects on an electron spin by its surrounding nuclear spins \cite{PhysRevLett.118.177702}.
 In the following, we shall assume that $\epsilon_a(t), \epsilon_d(t)$ are mutually independent stationary Gaussian processes with zero means. Under this  assumption, each noise is fully characterized in terms of its own power spectral density  $S_\mu(\omega) = \int_{-\infty}^{\infty} dt e^{-i \omega t} \langle \epsilon_\mu(0)\epsilon_\mu(t) \rangle, \mu \in \{a,d\}$. 
For practical applications, $S_\mu(\omega)$ will be determined from noise spectroscopy measurements in real experiments \cite{PhysRevLett.107.170504,PhysRevLett.107.230501}. 

Now, we briefly describe the inverse geometric optimization technique  \cite{PhysRevLett.111.050404,PhysRevLett.125.250403}. The procedure starts with a parameterization of the noise-free evolution. Let $U_0(t)$ be the solution to the time-dependent Schr{\"o}dinger equation
$\dot U_0(t) = -i H_0(t) U_0(t)$, where $H_0(t)$ is as shown in Eq. (\ref{eq1}) with $\epsilon_a,\epsilon_d=0$. 
We parameterize $U_0(t)$ based on $ZY{\!}Z$ decomposition, that is,  an arbitrary single-qubit unitary operator can be written as {  $ \exp(i \beta)R_z(\varphi) R_y(\theta) R_z(\gamma)$, for some real numbers $\beta,\varphi, \gamma \in [-\pi,\pi)$ and $\theta \in  [-\pi,\pi]$ \cite{NC10}.  In our problem here, $\beta = 0$ because  $H_0$ is traceless. }Hence, we have 
\begin{equation}
	U_0(t)  =  \left[ 
\begin{matrix}
 \cos({\theta}/{2}) e^{-i \varphi/2} e^{-i\gamma/2}   & -\sin(\theta/2) e^{-i \varphi/2} e^{i\gamma/2} \\
\sin({\theta}/{2}) e^{i \varphi/2} e^{-i\gamma/2}  &   \cos(\theta/2) e^{i \varphi/2} e^{i\gamma/2}
\end{matrix}
\right]. \nonumber
\end{equation} 
As such, the Schr{\"o}dinger equation is rewritten as
\begin{subequations}
\label{eq:qubit-par}	
\begin{align}
	\dot \theta & = \Omega \sin(\phi - \varphi),   \label{eq:theta} \\
	\dot \varphi & = -\Omega\cos(\phi - \varphi)   \cot \theta, \label{eq:varphi} \\
	\dot \gamma   & = \Omega\cos(\phi - \varphi) /\sin \theta.  \label{eq:gamma}
\end{align}
\end{subequations}
We perform optimization over these dynamical angular variables in order to find an evolution trajectory that has the property  of dynamically correcting errors on itself. 
In this geometric formulation of the control problem, the optimization objective consists of control target, robustness requirement, boundary conditions and certain practical considerations such as bounded control amplitude, all  expressed in terms of $\theta, \varphi$ and $\gamma$.   Once a robust evolution trajectory specified by the three angular variables is obtained, we can determine the control field by evaluating the inversion of Eq. (\ref{eq:qubit-par}), i.e.,
$
	\Omega  = \sqrt{{\dot{\theta}}^2 + {\dot \gamma}^2 \sin^2 \theta}, 
	\phi  = \arcsin (\dot \theta/\Omega) + \varphi
$.

\emph{Quantum gate and quantum state transfer.---}We first consider the control target of implementing a desired quantum gate or quantum state transfer. The key step is to effect the transformation operator to the toggling frame defined by  $U_{\epsilon_a,\epsilon_d}(t) = U_0(t)U_\text{tog}(t)$, where $U_{\epsilon_a,\epsilon_d}(t)$ represents the propagator in the presence of the noises. Through Dyson perturbative expansion \cite{PhysRev.75.486}, there is 
$U_\text{tog}(t) 
	   =  \mathds{1} -\sum_{\mu=a,d} [ 
		i \int_0^t dt_1 \epsilon_\mu(t_1) \widetilde E_\mu(t_1)+\int_0^t dt_1 \int_0^{t_1} dt_2 \epsilon_\mu(t_1)\epsilon_\mu(t_2) \widetilde E_\mu(t_1) \widetilde E_\mu(t_2) + \cdots ] 
$
with $\mathds{1}$ the identity operator and
$\widetilde E_\mu(t) = U_0^\dag(t) E_\mu U_0(t), \mu \in \{a,d\}$.  Substitute into the parameterized $U_0(t)$, we  obtain  $\widetilde E_{a,x}(t) = [\dot \theta \sin \gamma + (\dot \gamma \sin 2\theta \cos \gamma)/2]{\sigma_x}/{2}$, $\widetilde E_{a,y}(t) = [\dot \theta \cos \gamma - (\dot \gamma \sin 2 \theta \sin \gamma)/2]{\sigma_y}/{2}$, $\widetilde E_{a,z}(t) =(\dot \gamma \sin^2\theta){\sigma_z}/{2}$;
$\widetilde E_{d,x}(t) =  (- \sin \theta \cos \gamma){\sigma_x}/{2}$, $\widetilde E_{d,y}(t) = (\sin \theta \sin\gamma){\sigma_y}/{2}$, $\widetilde E_{d,z}(t) =(\cos\theta){\sigma_z}/{2}$.  These formulas are then to be substituted into the Dyson series to evaluate the error terms.

\begin{figure}
\includegraphics[width=\linewidth]{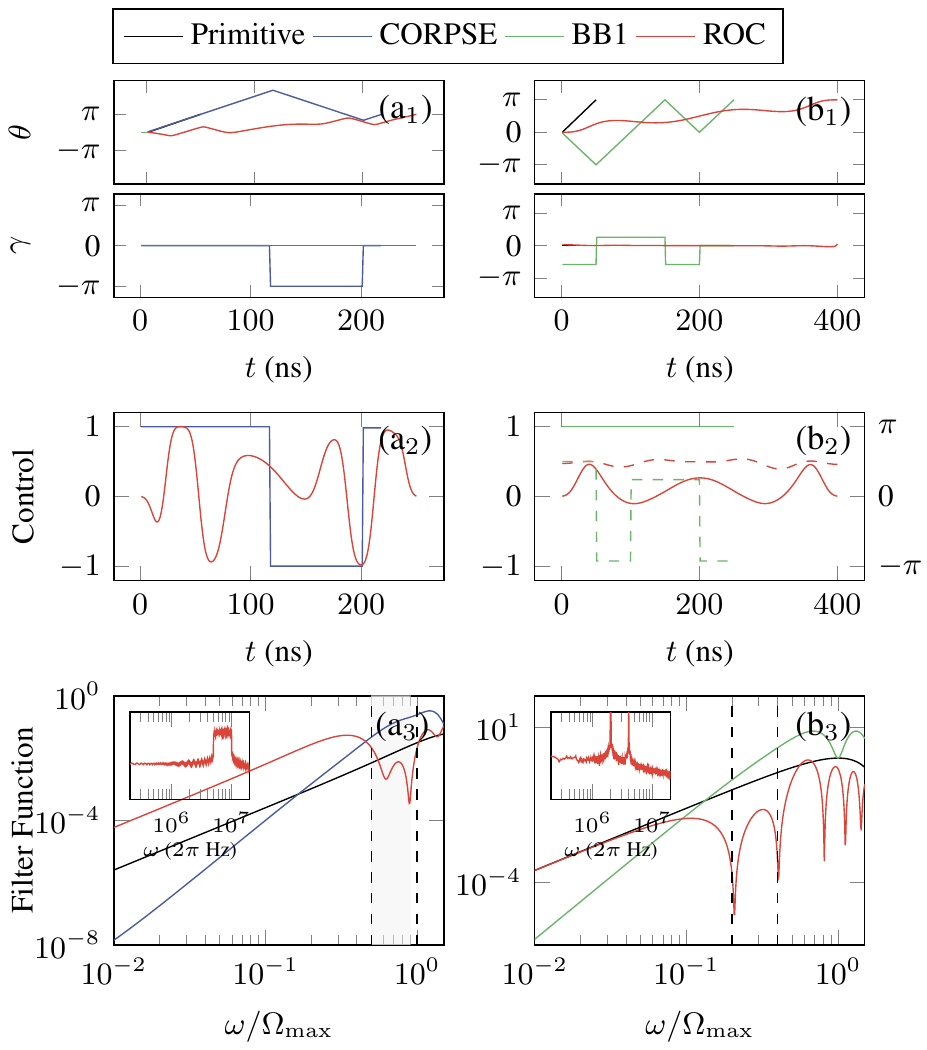}
\caption{   Geometric trajectories, control waveforms, FFs and noise spectra of different sequences for realizing a $\pi$ rotational gate  subject to time-dependent noise. (a$_1$)--(a$_3$) For detuning noise, the noise strength is set as $\sqrt{\langle \epsilon^2_d(0) \rangle} = 0.03 \Omega_{\max}$ with $\Omega_{\max}/(2\pi)=10^7~$Hz, and the noise spectrum is ohmic.
 (b$_1$)--(b$_3$) For amplitude noise, its strength is $\sqrt{\langle \epsilon^2_a(0) \rangle} = 0.03 $, and the noise spectrum is two Lorentzian peaks added on top of $1/f$ background ($\lambda_1=\lambda_2=100~$Hz, $\kappa=1, A_1=A_2=1,B=0.05$). The solid lines in (a$_2$) and (b$_2$) represent the pulse amplitudes in the unit of $\Omega_{\max}$, and the dashed lines in (b$_2$) are the pulse phases depicted by axis on the right.
 It can be seen that ROC FFs are suppressed at the characteristic frequencies, implying better noise filtering capability. Meanwhile, ROC control waveforms and geometric trajectories are much smoother.}
\label{fig2}
\end{figure}

For the  quantum gate   problem, we are given a target gate $\overline U$ and intend to find a robust implementing pulse.
Suppose that the ideal evolution at time $T$ satisfies $U_0(T) = \overline U$, then for a single realization of $\epsilon_a(t)$ and $\epsilon_d(t)$, the   gate fidelity reads $F = \left|\operatorname{Tr} \left( \overline U^\dag  U_{\epsilon_a,\epsilon_d}(T) \right)\right|^2/4 = \left|\operatorname{Tr} \left(   U_\text{tog}(T) \right)\right|^2/4$. 
Taking the ensemble average of the noises  and transferring to the frequency domain, the average gate infidelity defined by $\mathcal{F}_\text{avg} = 1- \langle F \rangle$ can be estimated to the second order approximation by the filter-function formalism \cite{GSUB13,SupplementalMaterial}
\begin{equation}
\label{eq:qubit-gatefidelity}
\mathcal{F}_\text{avg}    \approx  \frac{1}{2\pi}\sum_{\mu=a,d \atop \alpha=x,y,z} \int_{-\infty}^\infty \frac{d\omega}{\omega^2} S_\mu(\omega) |R_{\mu,\alpha}(\omega)|^2, 
\end{equation}
in which $R_{\mu,\alpha }(\omega) = -i\omega\int_0^T dt \text{Tr}[\widetilde E_{\mu,\alpha} (t) \sigma_\alpha/2] e^{i\omega t}$, {  and $\sum_\alpha |R_{d,\alpha}(\omega)|^2/\omega^2,\sum_\alpha |R_{a,\alpha}(\omega)|^2/(\omega^2 \Omega_{\max}^2)$ are the so called filter functions.} This formula provides a simple  quantitative means to evaluate the performance of a  control protocol in the presence of time-dependent noises.  It is thus natural to take $\mathcal{F}_\text{avg}$ as our objective function. As a concrete example, we consider implementing a $\pi$ rotational gate  $\overline U = \exp(-i\pi \sigma_y/2)$. For this problem, at $t=0$,   $U_0(0)$ equals to the identity, corresponding to the initial conditions $\theta(0)=0$ and $\varphi(0) = - \gamma(0)$ (value not specified). The ending point conditions are $\theta(T)=\pi$ and $\varphi(T) = \gamma(T) $. The latter can be rewritten as a constraint for $\theta$ and $\gamma$  by noting that from Eqs. (\ref{eq:varphi}) and   (\ref{eq:gamma}) there is $\dot \varphi = -\dot \gamma\cos(\theta) $,    hence one  requires the condition $ \gamma(0)  + \gamma(T) +\int_0^T   \dot  \gamma \cos \theta dt  = 0$ to be satisfied.  
{   With the objective function and all the constraints, we search ROC pulses using  the gradient-based algorithm \cite{khaneja2005optimal}; see details in Supplemental Material \cite{SupplementalMaterial}.}


For the quantum state transfer problem, without loss of generality,  we suppose the initial state to be $|0\rangle$. The target is an arbitrary state   $|\overline \psi\rangle$ on the Bloch sphere. 
For one  realization of the noise $\epsilon_a(t)$ and $\epsilon_d(t)$, the state transfer fidelity  reads $F = |\langle \overline \psi  | U_{\epsilon_a,\epsilon_d}(T) | 0\rangle |^2 $.  Suppose the ideal evolution $U_0(T)$ implements the desired state transfer. Again, we turn into the toggling frame and get $F = |\langle 0 | U_\text{tog}(T) | 0\rangle |^2 $. Substitute into the perturbative expansion of $U_\text{tog}(T)$,  and take ensemble average   of the noise Hamiltonian, it can be derived that   the   average state infidelity is 
\begin{equation}
\label{eq:qubit-statefidelity}
	\mathcal{F}_\text{avg}   \approx
	\sum_{\mu=a,d \atop \alpha=x,y,z }\frac{1}{2\pi}\int_{-\infty}^{\infty} \frac{d\omega}{\omega^2} S_\mu(\omega) |P_{\mu,\alpha}(\omega)|^2,
\end{equation}
where we define $P_{\mu,\alpha} = -i \omega \int_0^T dt \langle 0 | \widetilde E_{\mu,\alpha}(t) |1\rangle e^{i\omega t}$, {   and $\sum_\alpha |P_{d,\alpha}(\omega)|^2/\omega^2, \sum_\alpha |P_{a,\alpha}(\omega)|^2/(\omega^2 \Omega_{\max}^2)$ are the filter functions \cite{SupplementalMaterial}.} Concretely, we consider preparing target state  $|\overline \psi\rangle = |1\rangle$ {   starting from $|0\rangle$. This converts to the conditions} $\theta(0) = 0$, $\theta(T)=\pi, \varphi(0) = \varphi(T) = 0$, and no requirement of $\gamma$ is involved. Moreover, the relation between   $\theta$ and $\gamma$, namely $\dot \varphi = -\dot \gamma\cos(\theta) $,     requires the condition $ \int_0^T   \dot  \gamma \cos \theta dt  = 0$ to be satisfied.  The optimization procedure is the same as that described in  the gate problem.

{   As demonstration, we show the numerical simulation results of implementing a $\pi$ rotational gate under realistic detuning or amplitude noise, as shown in in Fig. \ref{fig2}. We compare performances between primitive, typical composite pulses and robust optimal control pulses, for the same given noise spectrum. The primitive pulse is the elementary rectangular pulse of maximum Rabi frequency $\Omega_{\max}$, which corresponds to the time-minimal control $t_{\min} = 1/(2\Omega_{\max})$. For detuning noise, as shown in Figs. \ref{fig2}(a$_1$)--\ref{fig2}(a$_3$), we consider ohmic spectrum with sharp cut-offs, i.e., $S_d(\omega) \propto \omega, \omega \in [\omega_{\text{lc}},\omega_{\text{uc}}]$, which describes a spin suffering bosonic environment \cite{RevModPhys.59.1}. The composite pulse we chosen is CORPSE \cite{bando2012concatenated}, which is robust to detuning error to the first order.  For amplitude noise, as shown in Figs. \ref{fig2}(b$_1$)--\ref{fig2}(b$_3$), we examine a noise spectrum of several Lorentzian peaks added on top of a broad $1/f^\kappa$ background $S_a(\omega) \propto \sum_k A_k/(\lambda_k^2+(\omega -\omega_{0,k})^2)+ B/\omega^\kappa $, which describes the random fluctuations in superconducting flux terms \cite{yan2013rotating}. The composite pulse tested for this case is BB1 \cite{bando2012concatenated},  which is robust to amplitude error to the second order.
It can be seen that in each test example, the ROC filter function  has sharp dips at   the central frequencies of the imported noise spectrum, hence their frequency overlap  is significantly suppressed; see Figs. \ref{fig2}(a$_3$) and \ref{fig2}(b$_3$). This feature implies that ROC  has  better performance in mitigating time-dependent noises.  We can  verify this conclusion by   computing their fidelities as follows. We   calculate a single instance of noise perturbed evolution operator $U_{\epsilon_\alpha}(T)$ and a single value for fidelity, and then take average over $N = 150$ noise realizations. For detuning noise with ohmic spectrum centered in the range $[0.5 \Omega_{\max},\Omega_{\max}]$, as shown in the insert of Fig. \ref{fig2}(a$_3$), we obtain $\mathcal{F}_\text{avg}^\text{ROC}= 4 \times 10^{-4}$, while $\mathcal{F}_\text{avg}^\text{Primitive}=1 \times 10^{-3}$ and $\mathcal{F}_\text{avg}^\text{CORPSE}=9 \times 10^{-3} $. This result is consist with the conclusion that composite pulses are only robust to fluctuating noises up to as fast as around 10\% of the Rabi frequency \cite{PhysRevA.90.012316}, yet our ROC pulse can still function for high-frequency noise. For amplitude noise with Lorentzian peaks centered at $0.2 \Omega_{\max}$ and  $0.4 \Omega_{\max}$ (see the insert of Fig. \ref{fig2}(b$_3$)), we obtain $\mathcal{F}_\text{avg}^\text{Primitive}=2 \times 10^{-3}$ and $\mathcal{F}_\text{avg}^\text{BB1}=8 \times 10^{-3}$, while our ROC pulse can decrease the infidelity to $\mathcal{F}_\text{avg}^\text{ROC}= 4 \times 10^{-6}$. Another benefit of  ROC pulse is that its shape  can be made much smoother than CORPSE and BB1; see Figs. \ref{fig2}(a$_2$) and \ref{fig2}(b$_2$).   This is particularly favorable for experiments, as real pulse generators   have limited bandwidths.  
Accordingly, ROC produces smoother geometric evolution trajectories, as shown in  Figs. \ref{fig2}(a$_1$) and \ref{fig2}(b$_1$). }

More simulation results for state transfer from $|0\rangle$ to $|1\rangle$, for the case when detuning and amplitude noise are simultaneously present, for other types of realistic noise models, and for varied characteristic frequency positions of the tested noise  spectra are all  put  in  the  Supplemental Material \cite{SupplementalMaterial}. These results reveal that, in general, ROC pulses  offer fidelity improvement for almost an order of magnitude compared with composite pulses and primitive pulse,  and  meanwhile featuring smooth pulse  shapes and geometric  trajectories.

\emph{Resistance of $T_1,T_2$ relaxation.---} When  noises vary  fast such that the Markovian approximation is valid, the controlled system dynamics can be described by the Bloch equation \cite{jeener1982superoperators} $\dot x =(H_0(t)+\gamma_1 R_{T_1} + \gamma_2 R_{T_2})x$, where $x\equiv (1/2,x_1,x_2,x_3)^T$ is the vectorized representation of the system density matrix $\rho=\mathds{1}/2 + x_1 \sigma_x +x_2 \sigma_y + x_3 \sigma_3$ $(x_1^2+x_2^2+x_3^2 \leq 1/4 )$, {  $H_0(t)$ is the control Hamiltonian,} $\gamma_{1,2}=1/T_{1,2}$ are relaxation rates and 
\begin{align}
{R_{{T_1}}} 
&= \left( {\begin{array}{*{20}{c}}
0&0&0&0\\
0&0&0&0\\
0&0&0&0\\
2 &0&0&{ - 1}
\end{array}} \right),
{R_{{T_2}}} = \left( {\begin{array}{*{20}{c}}
0&0&0&0\\
0&{ - 1}&0&0\\
0&0&{ - 1}&0\\
0&0&0&0
\end{array}} \right)
\end{align}
are their corresponding operators. Relaxation is an irreversible process, hence it is usually thought that the best strategy to alleviate effects of relaxation is to make the operation time as small as possible. Therefore, the primitive pulse  sets a fundament limit hard to surpass by other pulses  \cite{PhysRevLett.125.250403}. Here, we study this issue using inverse geometric optimization, which works as follows. We first parameterize the relaxation-free  evolution with extended three-dimensional rotations {  ${R_z}(\delta ), {R_y}(\eta )$ and ${R_z}(\xi )$}, namely {  ${V_0}(t) = \left( {\begin{array}{*{20}{c}}
1&0\\
0&{{R_z}(\delta ){R_y}(\eta ){R_z}(\xi )}
\end{array}} \right)$ with $\delta,\xi \in [-\pi,\pi], \eta \in [0,\pi]$} \cite{SupplementalMaterial}.
Thus, the Bloch equation is rewritten as 
\begin{subequations}
\label{eq:bloch}
\begin{align}
\dot \xi &=\Omega \cos(\phi-\delta)/\sin\eta,\\
\dot \eta &=\Omega \sin(\phi-\delta) ,\\
\dot \delta &=-\Omega \cos(\phi-\delta)/\tan\eta.
\end{align}
\end{subequations}
The actual evolution is then transformed to the toggling frame for conveniently displaying the perturbation effects due to $T_1$ and $T_2$ relaxation, i.e., $V_{T_1,T_2}(t)=V_0(t) V_\text{tog}(t) \approx V_0(t)(\mathds{1}_4+\sum_k \int_0^t d t_1 \gamma_k \widetilde R_{T_k}(t_1)+\cdots)$, where $\mathds{1}_4$ is the 4-dimensional identity, $\widetilde R_{T_k}(t)=V_0^\dag(t)R_{T_k}V_0(t)$.

\begin{figure} 
\includegraphics[width=0.95\linewidth]{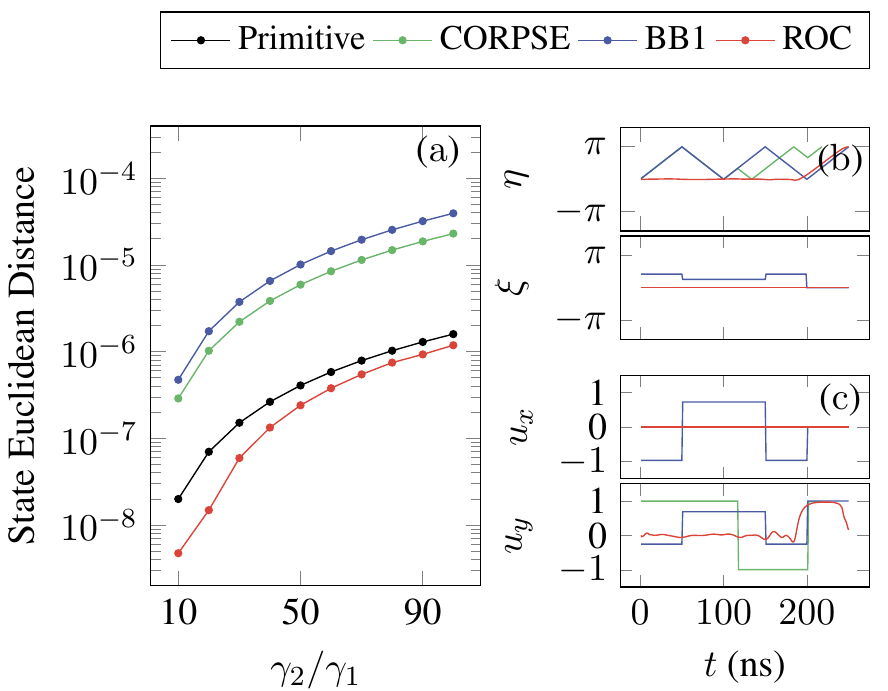}
\caption{Performance comparison of different sequences for realizing state transfer from the north pole to the south pole subject to both transverse and longitudinal relaxation. (a) State Euclidean distance vs  different relaxation parameters, where we set $\gamma_1=10^3~\text{s}^{-1}$. Specifically, we show geometric trajectories and control waveforms {  (in the unit of $\Omega_{\max}$)} for the case of $\gamma_2/\gamma_1=10$ in (b) and (c), respectively.
}
\label{openstate}
\end{figure}

Take the quantum state transfer problem as an example. Staring from state $x(0)$, the Euclidean distance between the actual state and the target state $\bar x$, defined by  $\mathcal{F}=|\bar x-V_{T_1,T_2}(T)x(0)|^2$, can be expressed in terms of the angular variables as follows \cite{SupplementalMaterial} 
\begin{equation}
	\mathcal{F}\approx  \int_0^T d{t}[\gamma_1^2(\cos\eta-2)^2 +\gamma_2^2 \sin^2\eta]/4.
\end{equation}
As a concrete example, we consider quantum state transfer from the north pole $x(0)=[1/2,0,0,1/2]^T$ to the south pole $\overline x=[1/2,0,0,-1/2]^T$.    This gives the  constraint  conditions $\eta(0)=0,\eta(T)=\pi$ and we set $\delta(0)=\delta(T)=0$, while no requirement of $\xi$ is involved. Besides, from Eqs. (\ref{eq:bloch}a) and (\ref{eq:bloch}c) we have $\dot \delta =-\dot \xi \cos\eta$, thus the condition $\int_0^T \dot \xi \cos\eta dt=0 $ should be satisfied. 
{   In our optimization, we also use the gradient-based algorithm to search robust ROC pulses.} Results are summarized in Fig. \ref{openstate}, where we consider solid-state spin defect system     with $\Omega_{\max}/(2\pi)=10^7~$Hz, and the relaxation parameters are typically chosen as $\gamma_1=10^3~\text{s}^{-1},\gamma_2/\gamma_1=10\sim 100$ \cite{wolfowicz2021quantum}. From Fig. \ref{openstate}(a), we find that typical composite pules, {   including CORPSE and  BB1,} can not resist relaxation, as they result in much larger errors compared with the primitive pulse. {  On the other hand, ROC pulses can improve up to four times compared with primitive pulse for all the tested relaxation parameters.} Meanwhile, the geometric trajectories and control waveforms for ROC pulses are smoother, as shown in Figs. \ref{openstate}(b) and \ref{openstate}(c), respectively.

{\emph{Discussion and outlook.---}The task of mitigating time-dependent noises  is   generally considered to be a thorny challenge and a long-term objective of quantum system engineering.  The robust control method presented here has a critical advantage of  flexibility as it is effective for a wide variety of noise environments, which is hence particularly applicable in reality since real experiments often involve
complicated noise spectrum.  Moreover, in the Markovian limit, the method is also effective  in improving the state transfer fidelity against transverse and longitudinal relaxation effects. We hope the control examples tested here or other possible applications can soon find their experimental verifications.

For future work, we can combine inverse geometric engineering with other robust optimal control methods. For example, the technique developed in Ref. \cite{HPZC19}, which expresses Dyson perturbative terms based on Van Loan's integral expression, provides a rather convenient means to evaluate the perturbative impacts of the noises. We can also apply analytic expression of the filter function derivatives \cite{PhysRevApplied.17.024006} to further improve the performance of our method, or attempt to derive exact analytical control fields \cite{PhysRevA.88.013818, PhysRevLett.109.060401}. In addition, the method presented here can be easily extended to handle other robust quantum control tasks, such as quantum sensing under time-dependent background noises \cite{titum2021optimal}. }

\emph{Acknowledgments}. We thank Ze Wu for helpful discussions. This work was supported by the National Natural Science Foundation of China (1212200199, 11975117, 92065111, 12075110, 11905099, 11875159, 11905111,  and U1801661), National Key Research and Development Program of China (2019YFA0308100), Guangdong Basic and Applied Basic Research Foundation (2019A1515011383 and  2021B1515020070),  Guangdong Provincial Key Laboratory (2019B121203002), Guangdong International Collaboration Program (2020A0505100001), Shenzhen Science and Technology Program (RCYX20200714114522109  and    KQTD20200820113010023),  China Postdoctoral Science Foundation (2021M691445), Science, Technology and Innovation Commission of Shenzhen Municipality (ZDSYS20190902092905285, KQTD20190929173815000 and JCYJ20200109140803865),  and Pengcheng Scholars, Guangdong Innovative and Entrepreneurial Research Team Program (2019ZT08C044).

\bibliography{manuscript.bib}

\newpage
\setcounter{equation}{0}
\renewcommand\theequation{S.\arabic{equation}} 

\setcounter{figure}{0}
 \renewcommand\thefigure{S\arabic{figure}}

\renewcommand{\bibnumfmt}[1]{#1.}

\onecolumngrid 
\vspace{100pt}
\begin{center}
{\large\bfseries Supplementary Material}	
\end{center}

\section{Inverse geometric optimization for single-qubit quantum system}

\subsection{Derivation of Eq. (2) of the Main Text: Evolution Parameterization}
An arbitrary single-qubit noise-free evolution can be parameterized by 
\begin{equation}
	U =\left[ \begin{matrix}
 \cos(\theta/2) e^{-i \varphi/2} e^{-i\gamma/2} & -\sin(\theta/2) e^{-i \varphi/2} e^{i\gamma/2} \\
\sin(\theta/2) e^{i \varphi/2} e^{-i\gamma/2}  &   \cos(\theta/2) e^{i \varphi/2} e^{i\gamma/2}
\end{matrix}
\right]. \nonumber
\end{equation}
Using this, the Schr{\"o}dinger equation can be rewritten as 
\begin{equation}
	\left[ \begin{matrix}
 \dot U_{11} & \dot U_{12} \\
\dot U_{21}  &  \dot U_{22}
\end{matrix}
\right] = \left[ \begin{matrix}
0 &   (-i u_x - u_y)/2 \\
(-i u_x + u_y)/2  &    0
\end{matrix}
\right] \left[ \begin{matrix}
 U_{11} &   U_{12} \\
  U_{21}  &    U_{22}
\end{matrix}
\right] \nonumber 
\end{equation}
which is
\begin{subequations}
\begin{align}
-\frac{\dot \theta}{2} \sin (\theta/2)   - \frac{i \dot \varphi}{2}\cos(\theta/2)   - \frac{i \dot \gamma}{2}\cos(\theta/2)   & = \frac{1}{2}(-i u_x - u_y)    \sin(\theta/2) e^{i \varphi},  \label{g11} \\
 - \frac{{\dot \theta }}{2}\cos (\theta /2) + \frac{{i\dot \varphi }}{2}\sin (\theta /2) - \frac{{i\dot \gamma }}{2}\sin (\theta /2){\rm{ }} &= \frac{1}{2}( - i{u_x} - {u_y})\cos (\theta /2){e^{i\varphi }}, \label{g12}\\
  \frac{\dot \theta}{2} \cos (\theta/2)   + \frac{i \dot \varphi}{2}\sin(\theta/2)   - \frac{i \dot \gamma}{2}\sin(\theta/2)   & = \frac{1}{2} (-i u_x + u_y)    \cos(\theta/2)  e^{-i \varphi}, \label{g21} \\
- \frac{{\dot \theta }}{2}\sin (\theta /2) + \frac{{i\dot \varphi }}{2}\cos (\theta /2) + \frac{{i\dot \gamma }}{2}\cos (\theta /2) &=  \frac{1}{2}( i{u_x} - {u_y})\sin (\theta /2){e^{ - i\varphi }}. \label{g22}
\end{align}
\end{subequations}
From these equations, we can obtain 
\begin{subequations}
	\begin{align}
\dot \theta & = -u_x \sin \varphi + u_y \cos \varphi= \Omega \sin(\phi - \varphi),  \\
	\dot \varphi & = -(u_x \cos \varphi + u_y \sin \varphi )   \cot(\theta) =-\Omega\cos(\phi - \varphi)   \cot \theta,  \\
	\dot \gamma  & = (u_x \cos \varphi + u_y \sin \varphi )/\sin(\theta) =\Omega\cos(\phi - \varphi) /\sin \theta. 
\end{align}
\end{subequations}
Once a robust evolution trajectory is obtained, we can determine the control fields by 
\begin{equation}
	\Omega(t)  = \sqrt{{\dot{\theta}}^2 + {\dot \gamma}^2 \sin^2 \theta}, 
	\phi(t)  = \arcsin (\dot \theta/\Omega) + \varphi.
\end{equation}

\subsection{Derivation of Eq. (3) of the Main Text: Average Gate Infidelity}
For one realization of noise $\epsilon_a(t)$ and $\epsilon_d(t)$, the real gate fidelity is
\begin{equation}
	F = \left|\operatorname{Tr} \left( \overline U^\dag  U_{\epsilon_a,\epsilon_d}(T) \right)\right|^2/4 = \left|\operatorname{Tr} \left( U_\text{tog}(T) \right)\right|^2/4. 
\end{equation}
It is convenient to write ${U_{{\rm{tog}}}}(T) = {\exp\{ - i \sum\limits_{\mu  = a,d} \int_0^T  {{\varepsilon _\mu }(t){{\tilde E}_\mu }(t)} dt \}} \equiv e^{{ - i \mathbf{a}\cdot \boldsymbol{\sigma}/2 }}= \mathds{1} \cos(a/2)-i \sin(a/2) \mathbf{a}\cdot \boldsymbol{\sigma}/a $, where $\boldsymbol{\sigma}=(\sigma_x,\sigma_y,\sigma_z)$, thus $F=[1+\cos(a)]/2$. For the first-order approximation, we obtain $F\approx 1-a^2/4$. Take the ensemble average of the noise, we get the average gate infidelity 

\begin{align}
	\mathcal{F}_\text{avg}
	&=1-\langle F \rangle \approx  \sum_{\mu=a,d} \left[  \int_0^T d {t_1}\int_0^T d {t_2} \epsilon_\mu(t_1)\epsilon_\mu(t_2)  \sum_{\alpha=x,y,z} \text{Tr}[\widetilde E_{\mu,\alpha} (t_1) \sigma_\alpha/2] \text{Tr}[\widetilde E_{\mu,\alpha} (t_2)\sigma_\alpha/2] \right] \nonumber \\
	&=\frac{1}{2\pi}\sum_{\mu=a,d \atop \alpha=x,y,z} \int_{-\infty}^\infty \frac{d\omega}{\omega^2} S_\mu(\omega) |R_{\mu,\alpha}(\omega)|^2.
\end{align}
where $R_{\mu,\alpha }(\omega) = -i\omega\int_0^T dt \text{Tr}[\widetilde E_{\mu,\alpha} (t) \sigma_\alpha/2]e^{i\omega t}$.

\subsection{Derivation of Eq. (4) of the Main Text: Average State Transfer Infidelity}
For one realization of noise $\epsilon_a(t)$ and $\epsilon_d(t)$, the real state fidelity is
\begin{align}
	F    = {} & |\langle \overline \psi  | U_{\epsilon_a,\epsilon_d}(T) | 0\rangle |^2 =  {}  |\langle 0 |   U_\text{tog}(T) | 0\rangle |^2  \nonumber \\
	  = {} & |\langle 0 | (\mathds{1} - \sum_{\mu=a,d}[i \int_0^T dt_1 \epsilon_\mu(t_1) \widetilde E_\mu(t_1) +  \int_0^T dt_1 \int_0^{t_1} dt_2 \epsilon_\mu(t_1)\epsilon_\mu(t_2) \widetilde E_\mu(t_1) \widetilde E_\mu(t_2) + \cdots])| 0\rangle |^2 \nonumber \\
	  \approx {} & 1 + \sum_{\mu=a,d}\int_0^T dt_1 \int_0^T dt_2 \epsilon_\mu(t_1)\epsilon_\mu(t_2)\langle 0 | \widetilde E_\mu(t_1) | 0\rangle \langle 0| \widetilde E_\mu(t_2) | 0\rangle^*  \nonumber \\
	{} &  -  \sum_{\mu=a,d}\int_0^T dt_1 \int_0^{t_1} dt_2 \epsilon_\mu(t_1)\epsilon_\mu(t_2)\langle 0 | \widetilde E_\mu(t_1) \widetilde E_\mu(t_2) | 0\rangle 
	- \sum_{\mu=a,d} \int_0^T dt_1 \int_0^{t_1} dt_2 \epsilon_\mu(t_1)\epsilon_\mu(t_2)\langle 0 | \widetilde E_\mu(t_1) \widetilde E_\mu(t_2) | 0\rangle^* ,
\end{align}
where we omit the cross terms containing $\epsilon_{\mu_1}(t_1)\epsilon_{\mu_2}(t_2)$ with ${\mu_1} \neq {\mu_2}$. 
Insert $|0\rangle \langle 0| + |1\rangle\langle 1|$ into the expression, then
\begin{equation}
F \approx     1 - \sum_{\mu=a,d }\int_0^T dt_1 \int_0^{T} dt_2 \epsilon_\mu(t_1)\epsilon_\mu(t_2)\langle 0 | \widetilde E_\mu(t_1) | 1\rangle \langle 1| \widetilde E_\mu(t_2) | 0\rangle.   	
\end{equation}	
Thus the average state infidelity is 
\begin{equation}
	\mathcal{F}_\text{avg} = 1- \langle F\rangle \approx
	\sum_{\mu=a,d \atop \alpha=x,y,z }\frac{1}{2\pi}\int_{-\infty}^{\infty} \frac{d\omega}{\omega^2} S_\mu(\omega) |P_{\mu,\alpha}(\omega)|^2,
\end{equation}
where $P_{\mu,\alpha} = -i \omega \int_0^T dt \langle 0 | \widetilde E_{\mu,\alpha}(t) |1\rangle e^{i\omega t}$.

\subsection{Gradient-based Optimization}
To search robust trajectory, we apply gradient-based optimization. 
{  In our optimization, we discretize   $\theta(t)$ and $\gamma(t)$ as $M$-slice sequences $\theta[1],...,\theta[M]$ and $\gamma[1],...,\gamma[M]$, respectively, with the  time length of each slice  $\tau = T/M$.}
For quantum gate, the derivative  of the average gate infidelity function reads
\begin{align}
	\frac{\partial \mathcal{F}_\text{avg}}{\partial \chi[m]} 
	 &= \sum_{\mu=a,d \atop \alpha=x,y,z }\frac{1}{2\pi} \int\limits_{-\infty}^\infty \frac{d\omega}{\omega^2} S_\mu (\omega)  \operatorname{Re}\left\{ \frac{\partial R_{\mu,\alpha}(\omega)}{\partial \chi[m]} R^*_{\mu,\alpha}(\omega)  \right\} \\ \nonumber 
	 &= \sum_{\mu=a,d \atop \alpha=x,y,z }\frac{1}{2\pi} \int\limits_{-\infty}^\infty \frac{d\omega}{\omega^2} S_\mu(\omega)  \operatorname{Re}\left\{ -i\omega \tau e^{i\omega m \tau} \frac{\partial \text{Tr}[\widetilde E_{\mu, \alpha}[m]\frac{\sigma_\alpha}{2}]}{\partial \chi[m]} R^*_{\mu,\alpha}(\omega)  \right\},\chi=\theta,\gamma.
\end{align}
Similarly, for quantum state transfer, the derivative  of the average state infidelity function reads
\begin{align}
	\frac{\partial \mathcal{F}_\text{avg}}{\partial \chi[m]} 
	 &= \sum_{\mu=a,d \atop \alpha=x,y,z }\frac{1}{2\pi} \int\limits_{-\infty}^\infty \frac{d\omega}{\omega^2} S_\mu (\omega)  \operatorname{Re}\left\{ \frac{\partial P_{\mu,\alpha}(\omega)}{\partial \chi[m]} P^*_{\mu,\alpha}(\omega)  \right\} \\ \nonumber 
	 &= \sum_{\mu=a,d \atop \alpha=x,y,z }\frac{1}{2\pi} \int\limits_{-\infty}^\infty \frac{d\omega}{\omega^2} S_\mu(\omega)  \operatorname{Re}\left\{ -i\omega \tau e^{i\omega m \tau} \frac{\partial \text{Tr}[\widetilde E_{\mu, \alpha}[m]|1\rangle\langle0|]} {\partial \chi[m]} P^*_{\mu,\alpha}(\omega)  \right\},\chi=\theta,\gamma.
\end{align}
The above gradients are then used to update the trajectory by $\chi[m]\leftarrow \chi[m]+ l \frac{\partial \mathcal{F}_\text{avg}}{\partial \chi[m]} $, where $l$ is appropriate step size.

\begin{figure*}
\includegraphics[width=0.9\linewidth]{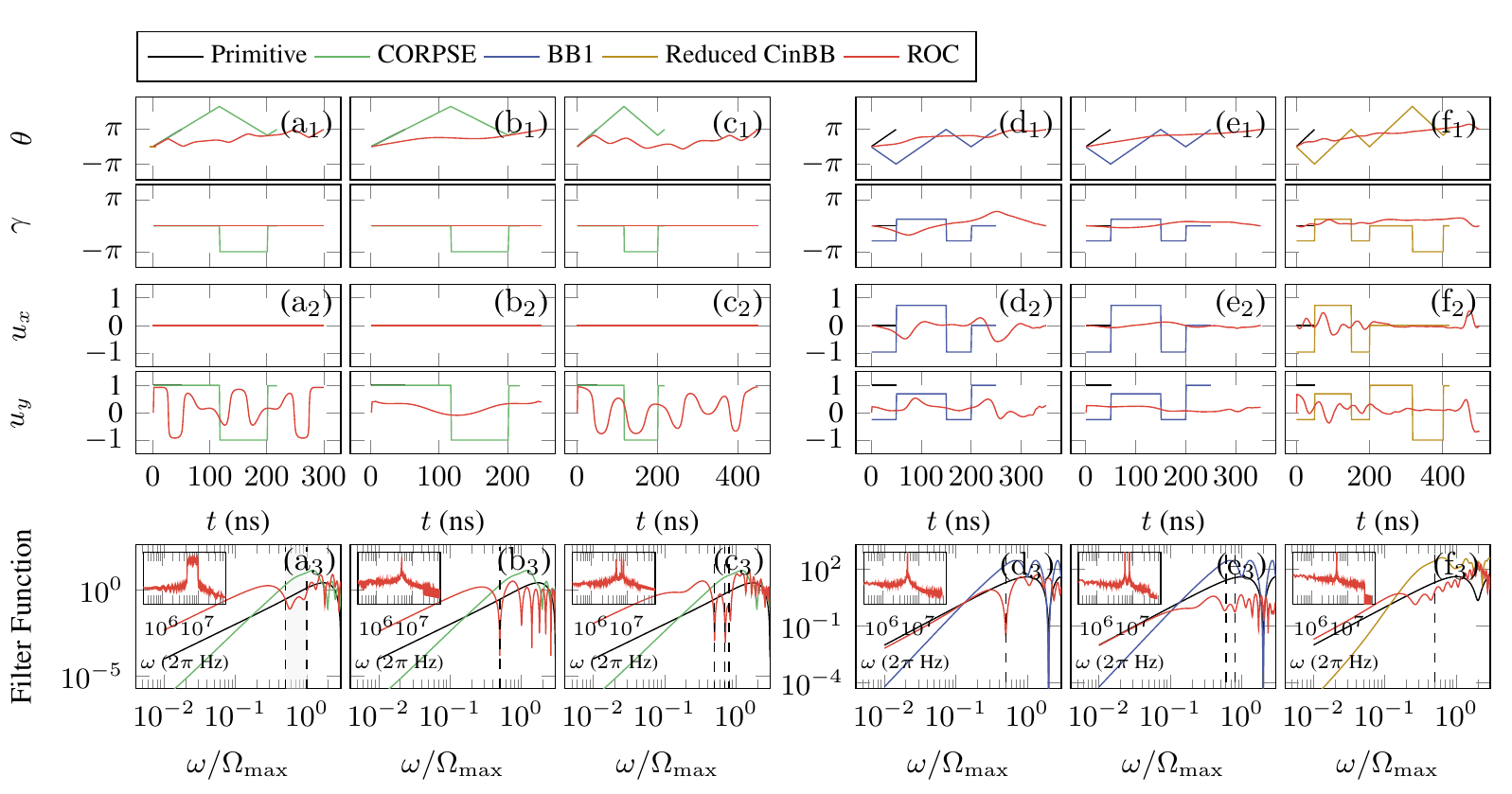}
\caption{Geometric trajectories, control waveforms (in the unit of $\Omega_{\max}$), FFs and noise spectrums  of  different sequences for realizing a $\pi$ rotation gate subject to high-frequency time-dependent ($\text{a}_1$)--($\text{a}_3$),($\text{b}_1$)--($\text{b}_3$),($\text{c}_1$)--($\text{c}_3$) detuning noise, ($\text{d}_1$)--($\text{d}_3$),($\text{e}_1$)--($\text{e}_3$) amplitude noise, or ($\text{f}_1$)--($\text{f}_3$) both. For detuning noise, the strength is $\sqrt{\langle \epsilon^2_d(0) \rangle} = 0.03 \Omega_{\max}$ with $\Omega_{\max}/(2\pi)=10^7~$Hz, and the noise spectrums are ohmic (insert in ($\text{a}_3$)), single Lorentzian (insert in ($\text{b}_3$), $\lambda =100~$Hz) , or multiple Lorentzian (insert in ($\text{c}_3$), $\lambda_1=\lambda_2=\lambda_3 =100~$Hz, $A_1=0.8,A_2=1.5,A_3=0.6$), respectively.
 For amplitude noise, the strength is $\sqrt{\langle \epsilon^2_a(0) \rangle} = 0.03 $, and the noise spectrums are a Gaussian peak added on top of $1/f$ background with a roll-off to white noise (insert in ($\text{d}_3$), $\sigma=5000~$Hz, $\kappa=1,A=1,B=0.05, \omega_\text{wc}=\Omega_{\max}$), or two Lorentzian peaks added on top of $1/f$ background (insert in ($\text{e}_3$), $\lambda_1=\lambda_2=100~$Hz, $\kappa=1, A_1=A_2=1,B=0.05$), respectively.
For simultaneous detuning noise and amplitude noise, the noise spectrums are both chosen as a single Lorentzian peak added on top of $1/f$ background (insert in ($\text{f}_3$), $\lambda=100~$Hz, $\kappa=1,A=1,B=0.05$) with the same strengths as above.}
\label{gate1}
\end{figure*}

\begin{figure*}
\includegraphics[width=0.9\linewidth]{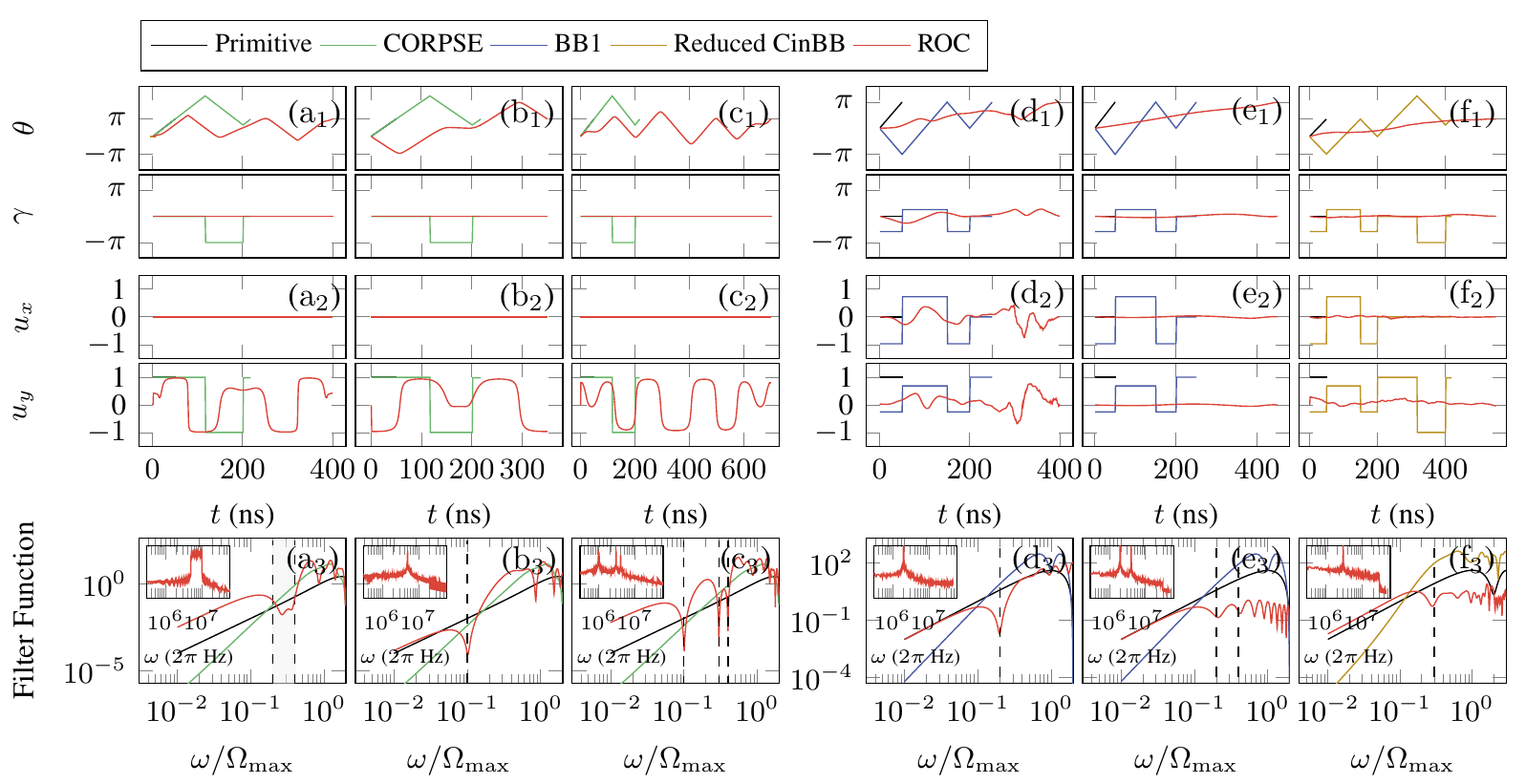}
\caption{Geometric trajectories, control waveforms, FFs and noise spectrums  of  different sequences for realizing a $\pi$ rotation gate subject to low-frequency time-dependent ($\text{a}_1$)--($\text{a}_3$),($\text{b}_1$)--($\text{b}_3$),($\text{c}_1$)--($\text{c}_3$) detuning noise, ($\text{d}_1$)--($\text{d}_3$),($\text{e}_1$)--($\text{e}_3$) amplitude noise, or ($\text{f}_1$)--($\text{f}_3$) both. For detuning noise, the noise spectrums are ohmic (insert in ($\text{a}_3$)), single Lorentzian (insert in ($\text{b}_3$), $\lambda =100~$Hz) , or multiple Lorentzian (insert in ($\text{c}_3$)), respectively.
 For amplitude noise, the noise spectrums are a Gaussian peak added on top of $1/f$ background with a roll-off to white noise (insert in ($\text{d}_3$)), or two Lorentzian peaks added on top of $1/f$ background (insert in ($\text{e}_3$)), respectively.
For simultaneous detuning noise and amplitude noise, the noise spectrums are both chosen as a single Lorentzian peak added on top of $1/f$ background (insert in ($\text{f}_3$)) with the same strengths as above. All the spectrum and control pulse parameters are the same with above case.}
\label{gate2}
\end{figure*}

\begin{figure*}
\includegraphics[width=0.9\linewidth]{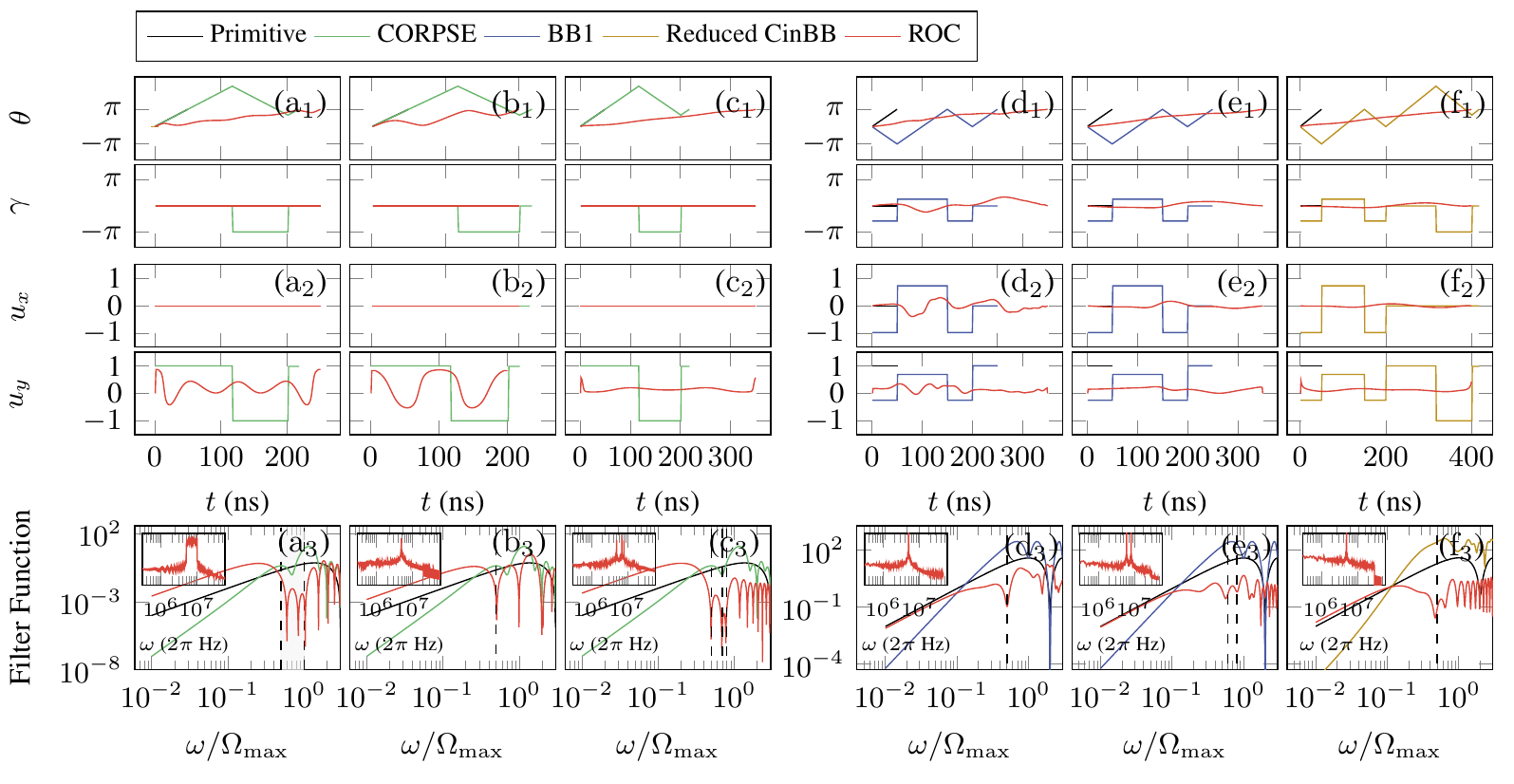}
\caption{ Geometric trajectories, control waveforms, FFs and noise spectrums of  different sequences for realizing state transfer from $|0\rangle$ to $|1\rangle$ subject to high-frequency time-dependent ($\text{a}_1$)--($\text{a}_3$),($\text{b}_1$)--($\text{b}_3$),($\text{c}_1$)--($\text{c}_3$) detuning noise, ($\text{d}_1$)--($\text{d}_3$),($\text{e}_1$)--($\text{e}_3$) amplitude noise, or ($\text{f}_1$)--($\text{f}_3$) both. For detuning noise, the noise spectrums are ohmic (insert in ($\text{a}_3$)), single Lorentzian (insert in ($\text{b}_3$)) , or multiple Lorentzian (insert in ($\text{c}_3$)), respectively.
 For amplitude noise, the noise spectrums are a Gaussian peak added on top of $1/f$ background with a roll-off to white noise (insert in ($\text{d}_3$)), or two Lorentzian peaks added on top of $1/f$ background (insert in ($\text{e}_3$)), respectively.
For simultaneous detuning noise and amplitude noise, the noise spectrums are both chosen as a single Lorentzian peak added on top of $1/f$ background (insert in ($\text{f}_3$)) with the same strengths as above. All the spectrum and control pulse parameters are the same with that in quantum gate case.}
\label{state1}
\end{figure*}

\begin{figure*}
\includegraphics[width=0.9\linewidth]{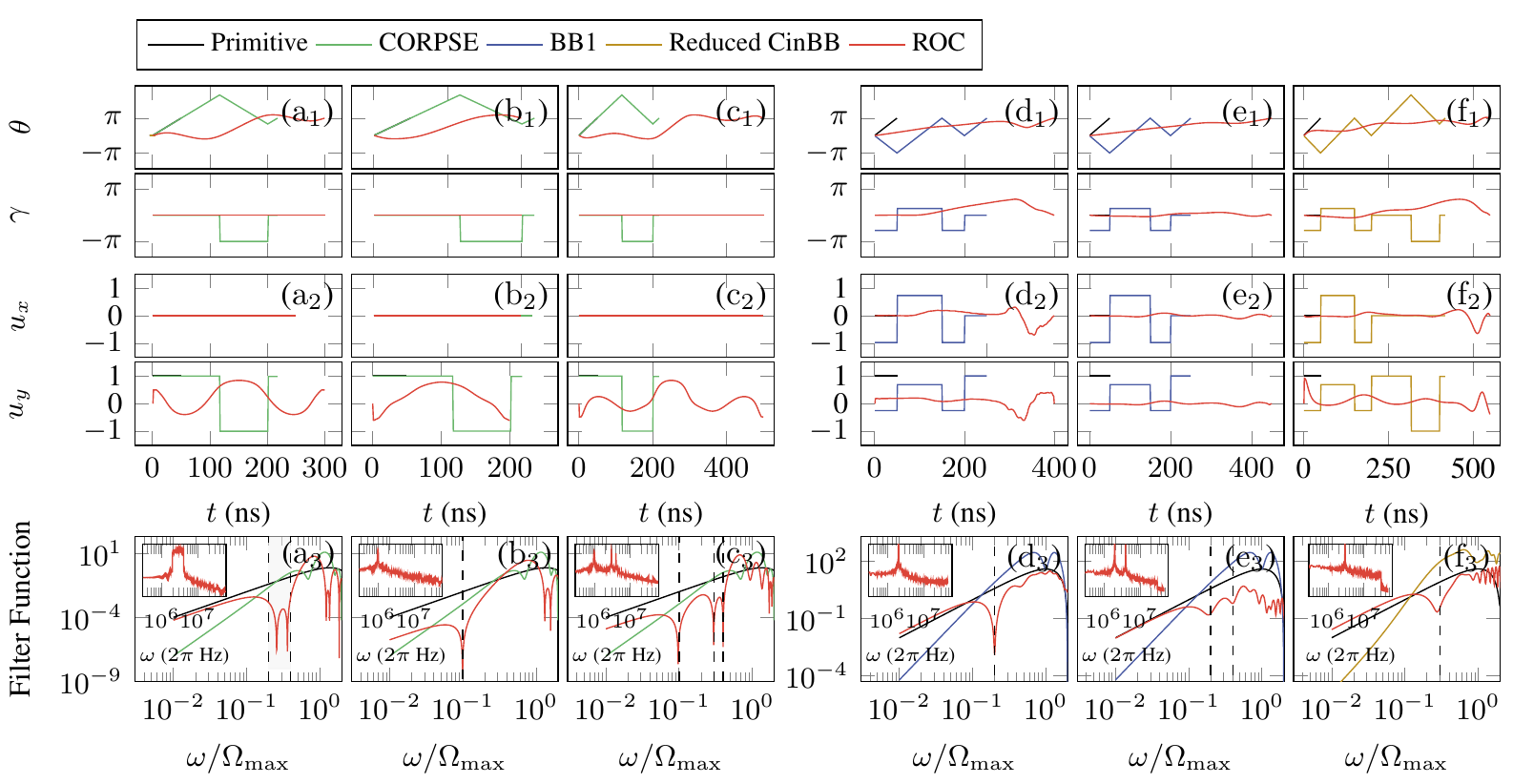}
\caption{ Geometric trajectories, control waveforms, FFs and noise spectrums of  different sequences for realizing state transfer from $|0\rangle$ to $|1\rangle$ subject to low-frequency time-dependent ($\text{a}_1$)--($\text{a}_3$),($\text{b}_1$)--($\text{b}_3$),($\text{c}_1$)--($\text{c}_3$) detuning noise, ($\text{d}_1$)--($\text{d}_3$),($\text{e}_1$)--($\text{e}_3$) amplitude noise, or ($\text{f}_1$)--($\text{f}_3$) both. For detuning noise, the noise spectrums are ohmic (insert in ($\text{a}_3$)), single Lorentzian (insert in ($\text{b}_3$)) , or multiple Lorentzian (insert in ($\text{c}_3$)), respectively.
 For amplitude noise, the noise spectrums are a Gaussian peak added on top of $1/f$ background with a roll-off to white noise (insert in ($\text{d}_3$)), or two Lorentzian peaks added on top of $1/f$ background (insert in ($\text{e}_3$)), respectively.
For simultaneous detuning noise and amplitude noise, the noise spectrums are both chosen as a single Lorentzian peak added on top of $1/f$ background (insert in ($\text{f}_3$)) with the same strengths as above. All the spectrum and control pulse parameters are the same with that in quantum gate case.}
\label{state2}
\end{figure*}

 {   
We test our robust control method for realizing a $\pi$ rotational gate or state transfer from $|0\rangle$ to $|1\rangle$ subject to various types of time-dependent noise, as shown in Figs. \ref{gate1}-\ref{gate2} and Figs. \ref{state1}-\ref{state2}, respectively. 
For detuning noise, we compare the performance between three sequences, namely primitive, CORPSE and ROC pulses, see the results in Figs. \ref{gate1}-\ref{state2}($\text{a}_1$)--($\text{a}_3$),($\text{b}_1$)--($\text{b}_3$),($\text{c}_1$)--($\text{c}_3$). Specifically, we consider (i) ohmic spectrum with sharp cut-offs, i.e., $S_d(\omega) \propto \omega, \omega \in [\omega_{\text{lc}},\omega_{\text{uc}}]$, which describes a spin suffering bosonic environment \cite{RevModPhys.59.1}; (ii) single Lorentzian spectrum $S_d(\omega) \propto 1/(\lambda^2+(\omega -\omega_0)^2)$ or multiple Lorentzian spectrum $S_d(\omega) \propto \sum_k A_k/(\lambda_k^2+(\omega -\omega_{0,k})^2)$ , which captures the solid-state spin environment of a spin bath as measured in, e.g., Refs. \cite{bar2012suppression,Hall15}. 
For amplitude noise, we compare the performance between primitive, BB1 and ROC pulses, as shown in Figs. \ref{gate1}-\ref{state2}($\text{d}_1$)--($\text{d}_3$),($\text{e}_1$)--($\text{e}_3$). Here we examine (i) a noise model corresponding to a strong and narrow Gaussian peak added on top of a broad $1/f^\kappa$ background with a roll-off to white noise, i.e., $S_a(\omega) \propto A \exp[-(\omega-\omega_0)^2/(2\sigma^2)]+ B/\omega^\kappa,\omega < \omega_\text{wc}$;   $S_a(\omega)=\text{const.}, \omega \geq \omega_\text{wc}$. This type of noise spectrum was observed, for example, in a silicon quantum dot spin qubit due to the imperfect control apparatus \cite{PhysRevApplied.10.044017}. (ii) several Lorentzian peaks added on top of a broad $1/f^\kappa$ background $S_a(\omega) \propto \sum_k A_k/(\lambda_k^2+(\omega -\omega_{0,k})^2)+ B/\omega^\kappa $, which describes the random fluctuations in superconducting flux terms \cite{yan2013rotating}. }
For simultaneous detuning noise and amplitude noise, for simplicity, we choose both of the noise spectrums as a single Lorentzian peak added on top of a broad $1/f$ background, i.e., $S_\mu(\omega) \propto A/(\lambda^2+(\omega -\omega_{0})^2)+ B/\omega$. We compare the performance between primitive, reduced CinBB and ROC pulses. Reduced CinBB \cite{bando2012concatenated} is a concatenated composite pulse for suppressing both of the detuning and amplitude noises. Results are summarized in Figs. \ref{gate1}-\ref{state2}($\text{f}_1$)--($\text{f}_3$). 
We also list explicitly the parameters of the noise power density spectrums and the tested control sequences in Table. \ref{table1}. All the simulations reveal that our  ROC method find high-quality, smooth and low-power robust pulses for resisting various realistic time-dependent noises.
In the main text, we demonstrate several typical results to show the effectiveness of our robust control method.

\begin{table*}
{\renewcommand\arraystretch{1.5}
\begin{tabular}{|p{2.9cm}<{\centering}|p{1.5cm}<{\centering}|p{0.9cm}<{\centering}|p{1.3cm}<{\centering}|p{1.3cm}<{\centering}|p{0.9cm}<{\centering}|p{1.3cm}<{\centering}|p{1.3cm}<{\centering}|p{0.9cm}<{\centering}|p{1.3cm}<{\centering}|p{0.9cm}<{\centering}|p{1.3cm}<{\centering}|}
\hline
 \multicolumn{2}{|c|}{Noise feature} & \multicolumn{3}{c|}{Primitive}&\multicolumn{3}{c|}{CORPSE/BB1/CinBB} &\multicolumn{4}{c|}{ROC}\\
\hline
  Noise spectrum type & $[\omega_{\text{lc}},\omega_{\text{uc}}]$ / $\omega_{0,k}$ ($*\Omega_{\max}$) & Pulse Length ($*T_{\text{P}}$)& Gate Infidelity  & State Infidelity &Pulse Length ($*T_{\text{P}}$) & Gate Infidelity &  State Infidelity & Pulse Length ($*T_{\text{P}}$) & Gate Infidelity & Pulse Length ($*T_{\text{P}}$)& State Infidelity \\
\hline
  \multirow{2}{*}{Ohmic} & $[0.5,1.0]$ & 1 & $1*10^{-3}$ & $6*10^{-4}$ & 4.3 & $9*10^{-3}$ & $2*10^{-3}$ & 6 & $4*10^{-4}$ & 5 & $1*10^{-5}$\\
  \cline{2-12}
  & $[0.2,0.4]$  & 1 & $1*10^{-3}$ & $9*10^{-4}$& 4.3 & $3*10^{-3}$ & $1*10^{-3}$& 8 &  $5*10^{-4}$ & 6& $1*10^{-5}$\\
 \hline
\multirow{2}{*}{Single Lorentizian} & 0.5 & 1 & $1*10^{-3}$ & $7*10^{-4}$ & 4.3 & $6*10^{-3}$ & $2*10^{-3}$ & 5 & $3*10^{-6}$& 4&  $2*10^{-5}$\\
\cline{2-12}
  & 0.1  & 1 & $9*10^{-4}$ & $9*10^{-4}$ & 4.3 & $4*10^{-4}$ & $1*10^{-4}$ & 7 & $3*10^{-5}$ & 4& $2*10^{-5}$\\
  \hline
\multirow{2}{*}{Multiple Lorentizian} & $0.5,0.7,0.8$ & 1 & $1*10^{-3}$ & $6*10^{-4}$& 4.3 & $9*10^{-3}$ & $8*10^{-4}$& 9 & $5*10^{-5}$& 7 & $4*10^{-6}$\\
\cline{2-12}
  & $0.1,0.3,0.4$ & 1 & $9*10^{-4}$ & $8*10^{-4}$ & 4.3 & $2*10^{-3}$ & $9*10^{-4}$ & 14 & $1*10^{-4}$& 10& $3*10^{-5}$\\
  \hline\hline
  Gaussian+$1/f$ & 0.5 & 1 & $2*10^{-3}$ & $2*10^{-3}$ & 5 &  $1*10^{-2}$ & $1*10^{-2}$ & 7 &  $3*10^{-5}$& 7& $2*10^{-5}$\\ 
  \cline{2-12}
  +white noise roll-off& 0.2 & 1 & $2*10^{-3}$ & $2*10^{-3}$ & 5 & $4*10^{-3}$ & $4*10^{-3}$& 8 & $3*10^{-5}$& 8& $7*10^{-5}$\\  
 \hline
Multiple Lorentizian & $0.6,0.8$  & 1 & $2*10^{-3}$ & $2*10^{-3}$ &5 & $1*10^{-2}$ & $1*10^{-2}$ &7 & $4*10^{-5}$& 7 & $6*10^{-5}$\\
\cline{2-12}
+$1/f$ & $0.2,0.4$ & 1 & $2*10^{-3}$ &$2*10^{-3}$ & 5 & $8*10^{-3}$ & $9*10^{-3}$ & 9 & $6*10^{-5}$& 9& $9*10^{-5}$\\  
\hline\hline
\multirow{2}{*}{Both: Lorentizian+$1/f$} 
& $0.5$ & 1 & $3*10^{-3}$ & $3*10^{-3}$  & 8.3 & $2*10^{-2}$ & $2*10^{-2}$  & 10 & $3*10^{-4}$& 8 & $3*10^{-4}$\\
\cline{2-12}
& $0.3$ & 1 & $3*10^{-3}$ & $3*10^{-3}$  & 8.3 & $3*10^{-2}$ & $4*10^{-2}$ & 11 & $3*10^{-4}$& 11 &$2*10^{-4}$\\
    \hline
\end{tabular}
}
\caption{Nose spectrum features and sequence parameters for realizing a $\pi$ rotational gate or state transfer from $|0\rangle$ to $|1\rangle$ subject to time-dependent detuning noise (the first sub table), amplitude noise (the second sub table) or both (the third sub table), respectively, where $\Omega_{\max}$ is the maximum rabi frequency, $T_\text{p}$ represents the length of primitive sequence.}\label{table1}
\end{table*}

{  \section{Inverse geometric optimization for two-level open quantum system}}
\subsection{Evolution Parameterization}
For an isolated spin-1/2 ensemble, the density matrix can be conventionally expressed as $\rho=\mathds{1}/2 + x_1 \sigma_x +x_2 \sigma_y + x_3 \sigma_3$. 
 When undergoing both transverse and longitudinal relaxation, and under Markovian approximation, the system dynamics under controls $u_x=\Omega(t)\cos(\phi(t)),u_y=\Omega(t)\sin(\phi(t))$ can be described by the Bloch equations
 \begin{equation}
 	\dot x =(H_0(t)+\gamma_1 R_{T_1} + \gamma_2 R_{T_2})x,
 \end{equation}
 where $x=(1/2,x_1,x_2,x_3)^T$, $\gamma_{1,2}=1/T_{1,2}$ are relaxation parameters and 
\begin{equation}
{H_0}(t) 
= \left( {\begin{array}{*{20}{c}}
0&0&0&0\\
0&0&-\Omega_0 &{u_y}\\
0&\Omega_0 &0&{-u_x}\\
0&{-u_y}&{u_x}&0
\end{array}} \right), 
{R_{{T_1}}} 
= \left( {\begin{array}{*{20}{c}}
0&0&0&0\\
0&0&0&0\\
0&0&0&0\\
{2{M_0}}&0&0&{ - 1}
\end{array}} \right),
{R_{{T_2}}} = \left( {\begin{array}{*{20}{c}}
0&0&0&0\\
0&{ - 1}&0&0\\
0&0&{ - 1}&0\\
0&0&0&0
\end{array}} \right).
\end{equation}
For simplicity, we set the off-resonance frequency $\Omega_0=0$ and the equilibrium state polarization $M_0=1$.

To apply inverse geometric optimization, we introduce the three-dimensional rotations 
\begin{equation}
R_{z}(\delta ) = \left( {\begin{array}{*{20}{c}}
{\cos \delta }&{ - \sin \delta }&0\\
{\sin \delta }&{\cos \delta }&0\\
0&0&1
\end{array}} \right),{R_y}(\eta ) = \left( {\begin{array}{*{20}{c}}
{\cos \eta }&0&{\sin \eta }\\
0&1&0\\
{ - \sin \eta }&0&{\cos \eta }
\end{array}} \right),{R_z}(\xi ) = \left( {\begin{array}{*{20}{c}}
{\cos \xi }&{ - \sin \xi }&0\\
{\sin \xi }&{\cos \xi }&0\\
0&0&1
\end{array}} \right).
\end{equation}
As such, the noise-free evolution of two-level open quantum system can the be parameterized by 
\begin{align}
{V_0}(t) 
&= \left( {\begin{array}{*{20}{c}}
1&0\\
0&{{R_z}(\delta ){R_y}(\eta ){R_z}(\xi )}
\end{array}} \right) \\ \nonumber
&=	\left( {\begin{array}{*{20}{c}}
 1 & 0 & 0 & 0 \\
 0 & \cos (\delta ) \cos (\eta ) \cos (\xi )-\sin (\delta ) \sin (\xi ) & -\sin (\delta ) \cos (\xi )-\cos (\delta ) \cos (\eta ) \sin (\xi ) & \cos (\delta ) \sin (\eta ) \\
 0 & \sin (\delta ) \cos (\eta ) \cos (\xi )+\cos (\delta ) \sin (\xi ) & \cos (\delta ) \cos (\xi )-\sin (\delta ) \cos (\eta ) \sin (\xi ) & \sin (\delta ) \sin (\eta ) \\
 0 & -\sin (\eta ) \cos (\xi ) & \sin (\eta ) \sin (\xi ) & \cos (\eta ) \\
\end{array}} \right),
\end{align}
thus the Bloch equation becomes
\begin{equation}
	{\dot V_0}(t) = {H_0}(t){V_0}(t),H_0(t)= \left( {\begin{array}{*{20}{c}}
0&0&0&0\\
0&0&0 &{ u_y}\\
0&0 &0&{-u_x}\\
0&{-u_y}&{u_x}&0
\end{array}} \right).
\end{equation}
This then gives
\begin{subequations}
	\begin{align}
		\dot \xi &= (u_x\cos\delta +u_y \sin\delta)/\sin\eta=\Omega \cos(\phi-\delta)/\sin\eta,\\
		 \dot \eta &= u_y\cos\delta-u_x\sin\delta=\Omega \sin(\phi-\delta) ,\\
		 \dot \delta &= -(u_x\cos\delta +u_y \sin\delta)/\tan\eta=-\Omega \cos(\phi-\delta)/\tan\eta.
	\end{align}
\end{subequations}
Once a robust evolution trajectory is obtained, we can determine the control fields by 
\begin{equation}
	\Omega (t) = \sqrt {{{\dot \eta }^2} + {{\dot \xi }^2}{{\sin }^2}\eta } ,\phi (t) =  \text{arcsin} (\dot \eta /\Omega )+\delta.
\end{equation}


\subsection{State Transfer Infidelity}
To characterize the distance between the actual state ${V_{{T_1},{T_2}}}(T)x(0)$ and the target state $\bar x$, we define 
\begin{align}
\mathcal{F} &= |\bar x - {V_{{T_1},{T_2}}}(T)x(0){|^2} = {[\bar x - {V_{{T_1},{T_2}}}(T)x(0)]^T}[\bar x - {V_{{T_1},{T_2}}}(T)x(0)] \\ \nonumber
&= {[\bar x - {V_0}(T){V_{{\rm{tog}}}}(T)x(0)]^T}[\bar x - {V_0}(T){V_{{\rm{tog}}}}(T)x(0)] \\ \nonumber
& \approx \left[ {V_0}(T)x(0) - {V_0}(T)(\mathds{1} + \sum\limits_{k = 1,2} {\int_0^T {d{t}} } {\gamma _k}{{\tilde R}_{{T_k}}}({t}))x(0) \right]^T \left[{V_0}(T)x(0) - {V_0}(T)(\mathds{1} + \sum\limits_{k = 1,2} {\int_0^T {d{t}} } {\gamma _k}{{\tilde R}_{{T_k}}}({t}))x(0) \right] \\ \nonumber
&=\left[\sum\limits_{k = 1,2} {\int_0^T {d{t}} } {\gamma _k}{{\tilde R}_{{T_k}}}({t})x(0) \right]^T \left[\sum\limits_{j = 1,2} {\int_0^T {d{t}} } {\gamma _j}{{\tilde R}_{{T_j}}}({t})x(0) \right] \\ \nonumber
&=\sum\limits_{k = 1,2}\left[{\int_0^T {d{t}} } {\gamma _k}{{\tilde R}_{{T_k}}}({t})x(0) \right]^T \left[{\int_0^T {d{t}} } {\gamma _k}{{\tilde R}_{{T_k}}}({t})x(0) \right],
\end{align}
where one should notice that $[{\tilde R}_{{T_k}}x(0)]^T[{\tilde R}_{{T_j}}x(0)] = 0$ when $k\neq j$.

\end{document}